\documentclass[12pt,english]{article}
\usepackage[latin1]{inputenc}
\usepackage{color}
\makeatletter

\newcommand{\be}{\begin{equation}}
\newcommand{\ee}{\end{equation}}
\newcommand{\bea}{\begin{eqnarray}}
\newcommand{\eea}{\end{eqnarray}}
\newcommand{\ba}{\begin{array}}
\newcommand{\ea}{\end{array}}

\def\bbox{{\,\lower0.9pt\vbox{\hrule \hbox{\vrule height 0.2 cm
\hskip 0.2 cm \vrule height 0.2 cm}\hrule}\,}}
\newcommand{\dsl}{\pa \kern-0.5em /}

\newcommand{\nn}{\nonumber \\}

\font\mybb=msbm10 at 12pt
\def\bb#1{\hbox{\mybb#1}}

\def\bR {\bb{R}}

\usepackage{babel}
\makeatother
\begin{document}
%%%%%%%%%%%%%%%% title page %%%%%%%%%%%%%%%%%%%%%%%%%%%%%%%%%%%%

%%%%%%%%%%%%%%%% title page %%%%%%%%%%%%%%%%%%%%%%%%%%%%%%%%%%
\begin{titlepage}

\begin{flushright}
%M5-BLG-rev7$_P$ \,
IFIC-08-33 \\ FTUV-08-2506 \\  DAMTP-2008-53
%$\hspace{2.1cm}{}$
\end{flushright}

\vfill

\begin{center}
\baselineskip=16pt {\Large\bf  Light-cone M5 and multiple M2-branes}

\vskip 2cm

{\large\bf Igor A. Bandos$^{\dagger, \ddagger , 1}$ and   Paul K.
Townsend$^{\star,2}$} \vskip 1cm {\small $^\dagger$ Departamento  de
F\'{\i}sica Teorica, Universidad de Valencia, \\ 46100, Burjassot,
Valencia,  Spain \\ } \vspace{12pt}
%{\small\it and \\}
{\small $^{\ddagger}$ Institute for Theoretical Physcs, \\
NSC Kharkov Institute of Physics \& Technology, \\
UA 61108,  Kharkov, Ukraine. \\
}
\vspace{12pt}

{\small
$^\star$
Department of Applied Mathematics and Theoretical Physics\\
Centre for Mathematical Sciences, University of Cambridge\\
Wilberforce Road, Cambridge, CB3 0WA, UK\\
}

\end{center}
\vfill

\par
\begin{center}
{\bf ABSTRACT}
\end{center}
\begin{quote}

We present the light-cone gauge fixed Lagrangian for the M5-brane; it has a residual
`exotic' gauge invariance with the group of 5-volume preserving  diffeomorphisms,
SDiff$_5$, as gauge group.  For an M5-brane of topology $\bR^2\times M_3$, for closed
3-manifold $M_3$,  we find an infinite tension limit that yields an $SO(8)$-invariant
$(1+2)$-dimensional field theory with `exotic'  SDiff$_3$ gauge invariance. We show that this
field theory is the Carrollian limit of the Nambu bracket realization of the `BLG'  model for
multiple M2-branes.

\vfill \vfill \vfill \vfill \vfill \hrule width 5.cm \vskip 2.mm {\small \noindent $^1$
bandos@ific.uv.es. Address after July 1: {\small $^{\dagger}$ Dept of Theor. Physics
and History of Science, The University of Basque Country (EHU/UPV), P.O. Box 644, 48080
Bilbao, Spain; supported by Ikerbasque (Basque Science Foundation).}
\\ \noindent $^2$ p.k.townsend@damtp.cam.ac.uk \\
}
\end{quote}
\end{titlepage}
%%%%%%%%%%%%%%%%%%%%%%%%%%%%%%%%%%%%%%

\section{Introduction}
\setcounter{equation}{0}

A $(1+2)$-dimensional relativistic gauge theory based on a Filippov 3-algebra
\cite{Filippov}  (see also \cite{Tak}) rather than a Lie algebra, was proposed recently
by Bagger and Lambert \cite{BL07}, and by Gustavsson \cite{G07}, as a model of multiple
M2-branes. The model has an $OSp(8|4)$ conformal symmetry \cite{Schwarz+08} as expected
for the infra-red fixed point  of the Yang-Mills-type gauge theory on coincident
$D2$-branes. The construction  requires a metric on the 3-algebra and if this metric is
positive definite then the structure constants of the 3-algebra define a
totally-antisymmetric fourth-rank tensor\footnote{Since the original  version of this
paper was posted on the archives,  it has been shown that `BLG-like' models can be
constructed from  a class of `generalized'  3-algebras for which this fourth-rank
tensor need not be totally antisymmetric \cite{Cherkis08}.} satisfying a `fundamental'
identity\footnote{Let $(B,C)$  be {\it anticommuting} variables taking values in a
Filippov $n$-algebra. Then the  fundamental identity  is equivalent to  $\{B, \dots,B,
\{C, \dots, C\}\} = n \{\{B,\dots,B,C\},C,\dots ,C\}$.}. When the structure constants
vanish one has a `trivial' 3-algebra and the model reduces to a free  theory for the
${\cal N}=8$ scalar multiplet, as expected for the conformal limit of a single planar
M2-brane. A non-trivial realization  based on the Lie algebra $so(4)$ was given by
Bagger and Lambert \cite{BL07}, and it appears to describe two coincident M2-branes on
an orbifold \cite{Lambert:2008et,Distler:2008mk}. It has since been shown that the only
other finite-dimensional realizations are direct sums of copies of this `$so(4)$-based'
algebra with trivial abelian 3-algebras \cite{Gauntlett:2008uf,Papadopoulos08}. Other
possibilities emerge when one allows for Lorentzian metrics on the 3-algebra
\cite{Russo08,BLGM5-M2D2} but these models have ghosts; we refer to some very recent
works for further discussion of this point \cite{JHS-BLG2,D2toD2}, and to
\cite{Eric+0608} for a supergravity perspective. Various other facets of
Bagger-Lambert-Gustavsson (BLG) models have been addressed in other papers; an
incomplete  list  can be found in \cite{BLG,BLG-Morozov,Krishnan:2008zm,G08}. In the
context of  the original  BLG  model, with positive definite metric, there remains one
other possibility: there is an infinite-dimensional realization of the 3-algebra in
terms of the Nambu bracket on a three-dimensional space
\cite{BLGM5,BLGM5-flux,BLGM5-M2D2}.  In this realization, the BLG model is essentially
an exotic gauge theory for the group of volume preserving diffeomorphisms of this
space, where by `exotic' we mean that the gauge theory is not of Yang-Mills type.

This is not the first occasion on which exotic gauge theories based
on volume-preserving diffeomorphisms have appeared. They also arise
from light-cone gauge fixing of  relativistic $p$-brane actions for
$p>2$;  these are `exotic' gauge theories with a group of $p$-volume preserving
diffeomorphisms, SDiff$_p$, as the gauge group  \cite{LCGp-branes}. This generalizes
the (dimensionally-reduced) Yang-Mills-type actions for $p=2$
where the Yang-Mills gauge group is a group of area-preserving
diffeomorphisms that may loosely be regarded as $SU(\infty)$
\cite{LCGM2-SU(inf)}. In particular, the light-cone gauge-fixed 10
dimensional (${\cal N}=1$) $5$-brane is an exotic gauge theory with
an SDiff$_5$ gauge group \cite{LCGp-branes}. Clearly, a similar
result should hold for the 11-dimensional M5-brane, and one purpose
of this paper is to present this SDiff$_5$-invariant  action.

Our starting point is the Hamiltonian M5-brane  action
for a general supergravity background \cite{BST98},  which  can be
deduced  from the Lorentz-covariant M5-brane action
\cite{blnpst,schw5} after  a `temporal gauge' choice  for
the PST-gauge invariance \cite{Pasti:1997gx}. One advantage of the
Hamiltonian form is that the passage to the light-cone gauge-fixed
theory is conceptually simpler. A further advantage, specific to the
M5-brane, is that  the non-linear self-duality of its worldvolume
3-form field strength $H=dA$ is very simply incorporated, off-shell,
by the disappearance from the action (excepting boundary terms) of the
time  components of the 2-form potential $A$. After partial fixing
of the worldvolume reparametrization invariance by the choice of
light-cone gauge, one is left with an $SO(9)$-invariant SDiff$_5$
exotic gauge theory.

 As is well-known, the flux of the 2-form potential on the M5-brane
may be interpreted as M2-branes `dissolved'  in the  M5-brane. Thus,
a single M5-brane may contain multiple M2-branes and is therefore a
promising starting point for a construction of the BLG model for
multiple-M2-branes. Another indication of this is that the
Nambu-bracket realization of the BLG theory introduces some
`internal' Riemannian 3-manifold $M_3$, so that the `total' space
dimension is $2+3=5$. In fact,  it has been  been proposed in recent
papers that the Nambu-bracket realization of the BLG model is {\it
equivalent} to the M5-brane action \cite{BLGM5,BLGM5-flux} (see
also \cite{CornM5mM2}). However, in the 11-dimensional Minkowski
vacuum of M-theory considered in  \cite{BLGM5,BLGM5-flux} and
here\footnote{A bosonic Minkowski background with constant supergravity 4-form is
gauge-equivalent to the M-theory vacuum.}, the symmetry algebra of the
M5-brane action  is an 11-dimensional  super-Poincar\'e algebra with tensor
charges \cite{Sorokin:1997ps}, so it would be remarkable if an $OSp(8|4)$
theory were to emerge.

Another purpose of this paper is to address this issue  from the
`opposite' direction: starting from the light-cone gauge fixed
M5-brane action, we consider an M5-brane of topology $\bR^2\times
M_3$ and then consider how the BLG theory, in its Nambu-bracket
realization,  might emerge from it.  It is natural to suppose that
the SDiff$_3$ gauge group of the Nambu-bracket BLG theory is a subgroup of the
SDiff$_5$ gauge group of the M5 theory, so we propose a partial
gauge fixing that identifies the $\bR^2$ coordinate with two of the
M5-brane coordinates.  This breaks the manifest $SO(9)$ invariance
to $SO(7)$, but we consider whether this could be enhanced to
$SO(8)$ in some limit. At the same time, we expect to find some
$(1+2)$-dimensional theory with an SDiff$_3$ residual gauge
group. One obvious way that this could happen is if  all fields
are assumed to be independent of position in $\bR^2$, but this amounts to a
double-dimensional reduction and it  yields an $SO(7)$-invariant
non-conformal 3-brane action on $M_3$,  rather than an
$SO(8)$-invariant  conformal theory on $\bR^2$.  Here,  we keep the
dependence  on all worldvolume coordinates, including the $\bR^2$
coordinates,  but we rescale the worldvolume fields by a power of
the M5-brane tension such that the rescaled fields
have the dimensions expected of a conformal $(1+2)$-dimensional theory,
and we also introduce rescaled dimensionless coordinates for $M_3$. We
then show that the infinite tension limit yields an SDiff$_3$ invariant gauge theory in
which $SO(7)$ is enhanced to $SO(8)$. In fact, the theory we get
this way differs from the BLG theory only in the absence of space
derivatives. Even though the fields depend on the $\bR^2$
coordinates, there are no derivatives with respect to them; we thus
find  a  ``Carrollian'' limit \cite{Carrollian,Bacry:1968zf} of the
BLG theory (see e.g. \cite{Gibbons:2003gb} for a recent discussion of this limit).

We will begin with a summary of some essential details of the
superspace geometry of  11-dimensional supergravity, and of the
M5-brane action,  and then proceed to our first result: the
light-cone gauge fixed action for an M5-brane in the
11-dimensional Minkowski vacuum of M-theory. We then consider M5-branes of topology
$\bR^2\times M_3$, partially gauge fix the SDiff$_5$ invariance, and
show how  a global $SO(8)$ and local SDiff$_3$ emerge
in a $T\to\infty$ limit that involves rescaling  fields and coordinates
by powers of $T$ to have the dimensions expected of a BLG theory.
We then summarize the results, explain their relation to the Carrollian limit of
 BLG theory, and speculate on possible extensions.

\section{Superspace and M5 preliminaries}
\setcounter{equation}{0}

An (on-shell) supergravity background is determined by the supervielbein one-form $E^A
=(E^{\underline{\alpha}},E^{\underline{a}})$ and the 3-form and 6-form potentials $C_3$
and $C_6$, subject to constraints on the torsion 2-form $T^A=DE^A$ and on the 4-form
and 7-form field strengths $R_4$ and $R_7$.  These constraints imply that the vector
component of the torsion 2-form  takes the form \be T^{\underline{a}}=
-iE^{\underline{\alpha}}\wedge E^{\underline{\beta}}\;
\Gamma^{\underline{a}}_{\underline{\alpha}\underline{\beta}}\, , \ee and that
\begin{eqnarray}
\label{R4:=dC=}
R_4 &=& dC_3 = E^{\underline{\alpha}}\wedge
E^{\underline{\beta}} \wedge
\bar{\Gamma}^{(2)}_{\underline{\alpha}\underline{\beta}}\,
+ {1\over 4!} E^{\underline{a}_4} \wedge \ldots \wedge E^{\underline{a}_1}
F_{\underline{a}_1\ldots \underline{a}_4 }\, , \\
R_7 &= & dC_6 +{1\over 2}C_3\wedge dC_3 = i
E^{\underline{\alpha}}\wedge E^{\underline{\beta}}  \wedge
\bar{\Gamma}^{(5)}_{\underline{\alpha}\underline{\beta}} + {1\over
7!} E^{\underline{a}_7} \wedge \ldots \wedge E^{\underline{a}_1}
F_{\underline{a}_1\ldots \underline{a}_7}\, \nonumber,
\end{eqnarray}
where
\be\label{bGn:=}
 \bar{\Gamma}^{(n)}:=  {1\over n!}
E^{\underline{a}_n} \wedge \ldots \wedge E^{\underline{a}_1}
{\Gamma}_{\underline{a}_1\ldots \underline{a}_n}  \, .
\ee

Let $Z^M$ be local coordinates for the 11-dimensional superspace, and let $\xi^m$ be
local coordinates for the M5-brane worldvolume. The embedding of the worldvolume in the
superspace is described by coordinate functions $Z^M(\xi)$ that define a map from the
worldvolume to the superspace. Differential forms on superspace may thereby be
pulled back to the worldvolume. We will use the same notation for  a superspace form
and its pullback as  the context should make it clear which is meant.  Thus, the
pullback of the supervielbein is \be E^A = d\xi^m E_m{}^A \, , \qquad E_m{}^A :=
\partial_m Z^M E_M{}^A\, , \ee and the induced worldvolume metric is $g_{mn}=
E_m{}^{\underline a}E_n^{\underline b}\, \eta_{\underline{a}\underline{b}}$, where
$\eta$ is the {\it mostly minus} Minkowski 11-metric.

As we will be using a Hamiltonian form of the M5-brane action, we set $\xi^m=(t,\sigma^i)$ ($i=1,2,3,4,5$) and we write
\be
E^A = dt E_t{}^A + d\sigma^i E_i{}^A\, .
\ee
The induced, {\it positive definite} metric on the 5-dimensional  `worldspace' is
 \be \label{gij=} ^5{}g_{ij}=-
 E_i{}^{\underline{a}}E_{j}{}^{\underline{b}} \eta_{\underline{a }\,\underline{b }}\, .
 \ee
We denote by $|^5g|$ the determinant of this metric.  Similarly, the pullback of the
3-form potential is
\begin{eqnarray}
\label{hC3:=} C_3={1\over 3!}d\xi^m\wedge d\xi^n\wedge d\xi^l
C_{lnm}:= {1\over 3!}dZ^M\wedge dZ^N\wedge
dZ^K {C}_{KLM}(Z)\, .
\end{eqnarray}
This is used to construct the worldvolume 3-form field-strength
$H= dA -C_3$ for the worldvolume 2-form potential $A$ of
the M5-brane.

We are now in a position to write down the Hamiltonian form of the
M5-brane action. More precisely, we choose  an intermediate form
that  requires only the  introduction of a Lorentz-vector momentum
variable $P_{\underline{a}}$, and a  time-space split; for example,
\begin{eqnarray}
A &=&  dt\wedge d\sigma^i A_{0i} + {1\over2}d\sigma^j\wedge d\sigma^i A_{ij}\nn
H &=&{1\over2} dt \wedge d\sigma^j \wedge d\sigma^i H_{0ij} + {1\over 6}d\sigma^k \wedge
d\sigma^j \wedge d\sigma^i H_{ijk} \, .
\end{eqnarray}
The feature of the action that results in  the non-linear self-duality of $H$  is a
constraint relating the  variables canonically conjugate to $A_{ij}$ to $H_{ijk}$
\cite{BST98}, and then $A_{0i}$ appears in the action only through a surface
term\footnote{In terms of the original covariant action \cite{blnpst}, this is a
consequence of the PST symmetry \cite{Pasti:1997gx}.}. The resulting
Lagrangian density is
\begin{eqnarray} \label{L(M5)=H}
 {\cal L}_{M5} &=&
P_{\underline{a}}E_t{}^{\underline{a}} +
 T\dot Z^M{\cal C}_M
- {T\over 8}  \varepsilon^{ijklm} \dot{A}_{ij}\
\partial_{k}A_{lm}   \\
&+& {s}^i\left(P_{\underline{a}}E_i{}^{\underline{a}} - T\sqrt{^5g}\,{\cal V}_i\right)
- {\ell\over 2} \left[
P_{\underline{a}}\eta^{\underline{a}\underline{b}}\,
P_{\underline{b}}- T^2\, |^5g|\,\left( 1 + {1\over 3!}H_{ijk}H^{ijk}
\right)\right]\, , \nonumber
\end{eqnarray}
where
\be
  \label{tj:=}
  {\cal V}_i:=  {1\over 4! \sqrt{|^5g|}}\varepsilon^{jklmn} H_{i jk}H_{lmn}\, ,
\ee
and
\begin{eqnarray}
\label{cCM:=} {\cal C}_M:=  {1\over 5!} \varepsilon^{ijklm}
C_{ijklmM} + {1\over 4!} \varepsilon^{ijklm}\left(C_{ijk}  +2H_{ijk}\right) C_{lmM} \, .
\end{eqnarray}
The variables $(\ell,s^i)$ are the `lapse' and `shift'  Lagrange multipliers for the
Hamiltonian constraint and worldspace diffeomorphism constraints, respectively. Note
that the `kinetic' term for $A$ is not manifestly gauge invariant but its gauge
variation is a total derivative.

In this paper, we consider only the 11-dimensional Minkowski vacuum, for which
\be
E^{\underline{a }}= dX^{\underline{a }}- i d\bar\Theta \Gamma^{\underline{a }}\Theta\, ,
\ee
where $\Theta$ is an $SO(1,10)$ Majorana spinor, so that
 \be \bar\Theta :=
\Theta^\dagger \Gamma^0 = \Theta^T C \, , \ee where $\Theta^T$ is
the transpose of $\Theta$ (viewed as a column vector), and $C$ is
the (unitary) antisymmetric charge conjugation matrix. In a Majorana
basis, the  (unitary) Dirac matrices are pure imaginary; for example
($\natural \equiv 10$)
\be \Gamma^0 = 1_{16} \otimes \sigma_2\, , \qquad
\Gamma^\natural = - 1_{16} \otimes i\sigma_1\, , \qquad \Gamma^I =
-\gamma^I\otimes i\sigma_3 \qquad (I=1,\dots,9),
\ee
where $1_{16}$ is the $16\times 16$ identity matrix, and $\gamma^I$ are the nine  $16\times 16$
{\it real} symmetric $SO(9)$ Dirac matrices,  satisfying $\{\gamma^I ,\gamma^J\}= 2\delta^{IJ} 1_{16}$.
In this basis we may choose $C= \Gamma^0$, so that $\Theta$ is a {\it real} 32-component spinor.

\section{Light-cone M5-brane}
\setcounter{equation}{0}

We choose coordinates such that the Minkowski 11-metric is
 \be
 ds^2_{11} = dX^{++} dX^{--} - dX^IdX^I\, , \qquad (I=1,\dots,9).
  \ee
The corresponding Dirac matrices, multiplied by the charge
conjugation matrix, are
 \be
C\Gamma^{++} = 2 \pmatrix{ 1_{16} & 0 \cr 0&0} \,
\qquad C\Gamma^{--} = 2 \pmatrix{0&0\cr 0 & 1_{16}}\,
, \qquad C\Gamma^I = \pmatrix{0&\gamma^I \cr \gamma^I &0}\, .
 \ee
In this basis,
 \be \Theta = \pmatrix{\theta_-\cr \theta_+}\, ,
 \ee
where $\theta_\pm$ are 16 component real $SO(9)$ spinors.

The light cone gauge is defined by
\be \label{LCG}
X^{++}=t ,   \qquad P_{--}  = -{T\over 4} \bar e\, , \qquad  \Gamma^{++}\Theta=0\, ,
\ee where $\bar e$ is the volume form for some (time-independent) `fiducial'
5-metric admitted by whatever topology we choose for the M5-brane, and the factor of $1/4$ is for later convenience. The constraint on $\Theta$ implies that $\theta_-=0$, as a result of which
 \begin{eqnarray}
\label{EinLCG} &&  E^{++}_\tau= 1 \; , \qquad E^{--}_\tau= \dot{X}{}^{--} -
2i \dot\theta_+ ^T\ \theta_+ \; , \qquad E^{J}_\tau= \dot{X}{}^{J}  \; ,
\qquad \nonumber
\\ \label{EjinLCG} &&  E^{++}_j= 0 \; , \qquad E^{--}_j=
\partial_j{X}{}^{--} - 2i \partial_j\theta_+^T\,  \theta_+\; , \qquad E^{J}_j=
\partial_j{X}{}^{J}\, ,
\end{eqnarray}
and the pullbacks of the superspace potentials are
\begin{eqnarray}
\label{C3lcg=} && C_3= -i d{X}^{++}\wedge dX^{J} \wedge
d\theta_+^T \wedge \gamma^{J}\theta_+ \; ,  \qquad
\\  \label{C6lcg=} && C_6={i\over 4!}\; d{X}^{++}\wedge dX^J
\wedge dX^K \wedge dX^L \wedge dX^M \wedge d\theta_+^T\wedge \gamma^{JKLM}
\theta_+\, .
\end{eqnarray}
Using
these results, we find that ${\cal C}_M = \delta_M^{++}{\cal
C}_{++}$, where
\begin{eqnarray} \label{cCM=lcg}
{\cal C}_{++} &=&  {i\over 24}\varepsilon^{ijklm}\partial_{i}
X^{I}\partial_{j} X^{J}\partial_{k} X^{K}
\partial_{l} X^{L}\; (\partial_{m} \theta_+^T \gamma^{IJKL}\theta_+) \nn
&&+\  {i\over 2}\varepsilon^{ijklm} \partial_{i} X^{J }\,
\left(\partial_{j}
\theta_+^T \gamma^{J}\theta_+\right) \partial_{k}A_{lm}\, .
\end{eqnarray}

The M5 Lagrangian density now reads
\begin{eqnarray}
\label{LlcgM5=3} {\cal L}_{M5} &=&
 \dot{X}^I P_I + {iT\over 2} \, \bar{e} \, \dot\theta_+^T\, \theta_+
 \,  - {T\over 8} \dot{A}_{ij}\varepsilon^{ijklm}\partial_{k}A_{lm}  - {T
|{^5g}|\over \bar{e}  }\, \left(1 + {1\over 3!}H_{ijk}H^{ijk} \right) \nonumber \\ &&
- \ {P_IP_I \over T\bar{e}  } + T{\cal C}_{++}  +  \bar e s^j K_j + {T\over 4} X^{--}
\partial_j (\bar{e}s^j)\; , \qquad
\end{eqnarray}
where
\begin{eqnarray}
\label{Kj=} \bar e K_i:=  \partial_i X^I P_I - \; {T \over 4!}\;
\varepsilon^{jklmn} H_{ijk}H_{lmn} + {iT\bar{e} \over 2 }
\;\partial_i \theta_+^T \theta_+ \, .
\end{eqnarray}
The variable $X^{--}$ is now a Lagrange multiplier imposing the constraint
 \be\label{VD}
\partial_i\left(\bar e s^i\right)=0\, .
 \ee
This condition is solved {\it locally} by
 \be \bar e s^i = \varepsilon^{ijklm} \partial_j \Sigma_{klm}\, ,
 \ee
where $\Sigma_{klm}$ is the unconstrained Lagrange multiplier for the constraint
$\partial_{[i}K_{j]}=0$, defined up to an obvious abelian gauge transformation.
It imposes the vanishing of  $\partial_{[i}K_{j]}$, which generates (via Poisson brackets)
the 5-volume-preserving diffeomorphisms of the canonical variables.

Rather than solve the constraint for $s^i$, we may proceed on the understanding that it
is constrained by (\ref{VD}). We may then rewrite the Lagrangian density as
\begin{eqnarray}
\label{covdivform} {\cal L}_{M5} &=& D_t X^I P_I + {iT\over 2} \, \bar{e} \, D_t\theta_+^T  \,
\theta_+ \,  - {T\over 4!}\varepsilon^{ijklm} \left(D_t A\right)_{ij} H_{klm} \nn
&&-\,  {P_IP_I \over T\bar{e} }   - {T |{^5g}|\over \bar{e}  }\, \left(1 + {1\over
3!}H_{ijk}H^{ijk} \right) + T{\cal C}_{++} \, ,
\end{eqnarray}
where $D_t$ is a covariant time derivative:
 \begin{eqnarray}\label{DtX:=}
D_t  X^I &:=& \dot X^I + s^j\partial_j X^I \, \qquad
D_t \theta_+ = \dot\theta_+ + s^j\partial_j \theta_+  \nn
\left(D_t A\right)_{ij} &:=& \dot A_{ij} + s^k H_{kij}\, , \qquad
D_tP_I = \dot P_I + \partial_j\left(s^j P_I\right)\, .
\end{eqnarray}
The infinitesimal SDiff$_5$ gauge transformations are
\begin{eqnarray}\label{infinitesimal}
\delta X^I &=& - \zeta^i \partial_i X^I \, , \qquad \delta\theta_+ = - \zeta^i \partial_i \theta_+ \, , \qquad
\delta P_I  = -\partial_i \left(\zeta^i P\right)\, , \nn
\delta A_{ij} &=& - \zeta^k\partial_k A_{ij} + 2 \partial_{[i}\zeta^k\,  A_{j]k} \, , \qquad
\delta s^i  = \dot\zeta^i + \left[s,\zeta\right]^i\, ,
\end{eqnarray}
where $[,]$ is the Lie bracket of  worldspace vector fields, and the vector parameter  $\zeta$ satisfies
\be
\partial_i \left(\bar e \zeta^i\right)=0\, .
\ee

Setting to zero all fermions and the gauge fields, for simplicity,
and eliminating $P_I$ as an auxiliary field, we arrive at an
SDiff$_5$-invariant Lagrangian density of the form
\be
{\cal  L} = {1\over4}T \bar e \left| D_t X \right|^2 - \bar e V\, .
\ee
The potential is
\be \label{VM5=}
V= {T\over \bar e^2} \left|{}^5g\right| = {T\over 5!} \sum_{I,J,K,L,M}
\left\{X^I,X^J,X^K,X^L,X^M\right\}^2\, ,
 \ee
where
 \be\label{NambuB5}
\left\{X^I,X^J,X^K,X^L,X^M\right\} := \bar e^{-1}\varepsilon^{ijklm}
\partial_i X^I \partial_j X^J \partial_k X^K\partial_l X^M\partial_l X^L\, ,
 \ee
which is a generalized Nambu bracket, itself a generalization of the Poisson bracket.
Note that we define the bracket with the inverse of the `fiducial' density $\bar e$ in order that
it map products of scalars to a scalar\footnote{The analogous analysis for the M2-brane leads to a
similar result but with a bilinear Poisson bracket instead of a multi-linear Nambu bracket.
The Poisson bracket may again  be defined  with a factor of  $\bar e^{-1}$, because this is consistent with the Jacobi identity, and it  {\it should}  be so defined in order that products of scalars get mapped to a scalar.  To see the necessity of this factor, it suffces to consider a spherical M2-brane
such that $X^2+Y^2+Z^2=1$; one finds that $\{X,Y\}_{PB}=Z$, and cyclic permutations,
only if $\bar e$ is the volume form on the unit sphere.}.

As a prelude to the  procedure considered in the remainder of the paper, we
now show how a rescaling of  the variables by appropriate powers of the tension $T$
allows all dependence on $T$ to be factored out. Specifically, we set
 \be\label{rescale}
 X^I = T^{-\nu} \tilde X^I\, , \qquad P_I = T^{1-\nu}\tilde P_I\, ,
\qquad A= T^{-\nu} \tilde A\, ,  \qquad \theta_+ = T^{-\nu} \tilde \theta_+\, ,
 \ee
for arbitrary real constant $\nu$.  The result is that
\be
I[X,P,A,\theta_+] = T^{1-2\nu}
\tilde I[\tilde X,\tilde P,\tilde A,\tilde\theta_+]\, ,
\ee
where $I$ the action functional with
$T$-dependent integrand $L_{M5}$ of (\ref{LlcgM5=3}) and $\tilde I$ is the same
functional but with $T=1$.  For $\nu=1/2$, the $T^{1-2\nu}$ prefactor  is unity and the
dimensions of  the variables become the standard dimensions  for fields in a
six-dimensional spacetime.

\section{Further gauge fixing and a hypertensile  limit}
\label{sec:hyper}
\setcounter{equation}{0}

We now suppose that the M5-brane has topology $\bR^2\times M_3$ for some compact 3-manifold $M_3$. This means that we may choose local tranverse space coordinates $X^I$, and local worldspace
coordinates $\sigma^i$, such that
\begin{eqnarray}\label{furtherGF}
X^I  &=& \left(X^{\hat I}, x^\alpha\right)\, ,  \qquad \hat I = 1,\cdots, 7 \, , \qquad \alpha=1,2 \; ,  \nonumber\\
\sigma^i &=&  \left(\sigma^{\dot\alpha}, x^\alpha\right)\, , \qquad  \dot\alpha=1,2,3 \, ,
\end{eqnarray}
where $x^{\alpha}$ are cartesian coordinates for $\bR^2$. The
fiducial worldspace density $\bar e$ should now be understood as a
worldvolume density on $M_3$, independent of time and the $\bR^2$
coordinates.  We may also rewrite the invariant worldspace
alternating tensor density:
\be
\varepsilon^{ijklm}  \to
\varepsilon^{\alpha\beta}\varepsilon^{\dot\alpha\dot\beta\dot\gamma}\, .
\ee
An implication
of our partial gauge choice $X^\alpha= x^\alpha$ is that the
manifest $SO(9)$ invariance is broken to a manifest $SO(7)$
invariance\footnote{There is still an $SO(9)$ invariance, of course, but it becomes part of the non-linearly realized $SO(1,10)$ invariance.}. Accordingly, we split the $SO(9)$ Dirac matrices into
reducible $SO(7)$ Dirac matrices $\gamma^{\hat I}$ and the two
matrices  $\gamma^\alpha$, which are reducible ($16\times 16$) Dirac
matrices for $\bR^2$. These matrices have the anti-commutators
 \be
\left[\gamma^{\hat I},\gamma^{\hat J}\right]_+ = \delta^{\hat I \hat
J} 1_{16}\, , \qquad \left[\gamma^\alpha,\gamma^\beta\right]_+ =
\delta^{\alpha\beta} 1_{16}\, .
 \ee

Having eliminated  the variables $X^\alpha$ by a gauge choice, we expect to be able to
express the conjugate variables $P_\alpha$ in terms of the remaining variables, and
this will eventually be done. However, we postpone this step as it can be done more
simply after we have settled other issues. One such issue is whether there is a
`hidden'  {\it linearly-realized} $SO(8)$ invariance, as suggested by the fact that
the  $\bR^2$ component of the worldvolume  2-form $A$ may be identified as an 8th
scalar. As we shall see, there is an `enhancement' of $SO(7)$ to $SO(8)$ but only in a
particular  infinite tension (hypertensile) limit that involves first
rescaling the fields and coordinates. Another question is the nature of the residual
gauge group. The problem is that $\tilde s^{\dot\alpha}$ does not satisfy its own
divergence-free condition, as would be expected for an SDiff$_3$ gauge theory; instead,
its divergence is related to the divergence of $s^\alpha$ by the constraint (\ref{VD}).
As we shall see, this problem is resolved in the hypertensile limit.

We first consider a rescaling of the fields of the type (\ref{rescale}).  Leaving aside the  two-form potential for the moment, this means that
 \be\label{Th=TPsi}
 X^{\hat I}
= T^{-\nu} \phi^{\hat I}\, , \qquad P_{\hat I} = T^{1-\nu}\pi_{\hat I}\, , \qquad
\theta_+ = T^{-\nu} \Psi\, ,
 \ee
for new variables $(\phi^{\hat I},\pi_{\hat I}, \Psi)$.  As a consequence of this
rescaling,
 \be\label{kineticXP} \dot X^I P_I + {iT\over 2} \, \bar{e} \, \dot
\theta_+^T\, \theta_+ = T^{1-2\nu} \left[ \dot \phi^{\hat I} \pi_{\hat I}  + {i\over 2}
\, \bar{e} \, \dot \Psi^T \Psi\right]\, . \ee Ultimately, we will choose
  \be
 \nu= 1/4
 \ee
because this yields dimensions for the rescaled fields that are  appropriate for a
conformal  theory in $(1+2)$ dimensions; recall that $T$ has mass dimension $[T]= 6$ in
fundamental units. This choice leads to an overall factor of $\sqrt{T}$ but this can be cancelled,
 in the  {\it action},  by a rescaling of the $M_3$ coordinates. This requirement fixes the rescaling of the $M_3$ coordinates for the choice $\nu=1/4$ but we will find it convenient to retain  $\nu$ as a free positive parameter on the understanding that  $\nu=1/4$ will be our ultimate choice. For other values of $\nu$ the rescaling of the $M_3$ coordinates can be fixed by requiring that all leading terms in the Lagrangian density for large $T$ appear with the same factor of $T^{1-2\nu}$. As we shall verify, this
 happens when the $M_3$ coordinates are rescaled such that
 \be \label{d=Td}
\partial_{\dot\alpha} = T^{2\nu/3} \tilde\partial_{\dot\alpha} \, .
 \ee
For  $\nu=1/4$ the rescaled coordinates are dimensionless and $d^3 \sigma = (1/\sqrt{T})d^3\tilde{\sigma}$, as required.

If we now define a new pair of variables $(\phi^8,\pi_8)$ by \be
\frac{1}{2}\varepsilon^{\alpha\beta} A_{\alpha\beta} = T^{-\nu}
\phi^8\, , \qquad
\frac{1}{3!}\varepsilon^{\dot\alpha\dot\beta\dot\gamma}H_{\dot\alpha\dot\beta\dot\gamma}
=- T^{-\nu} \pi_8\, , \ee and rescaled mixed  and $M_3$ components
of $A$ by
 \be\label{A=TtA} A_{\alpha\dot{\beta}} = T^{-4\nu/3}
\; b_{\alpha\dot{\beta}}\, ,  \qquad A_{\dot{\alpha}\dot{\beta}} =
T^{-2\nu/3}\; \tilde{A}_{\dot{\alpha}\dot{\beta}}\, ,
 \ee
then
 \be\label{dotAdA=}
- {T\over 8} \varepsilon^{ijklm}\dot{A}_{ij} \, \partial_{k}A_{lm} =
T^{1-2\nu}\left[ \dot{\phi}^8\pi_8 - \frac{1}{2}
\varepsilon^{\alpha\beta} \dot{b}_{\alpha\dot\alpha}
\varepsilon^{\dot\alpha\dot\beta\dot\gamma} \left(
\tilde\partial_{\dot\beta}b_{\beta\dot\gamma} - \partial_\beta
\tilde A_{\dot\beta\dot\gamma}\right)\right]\, . \ee
 The $\dot{\phi}^8\pi_8$ term provides an $SO(8)$ completion  of (\ref{kineticXP}).  The second term transforms into a total derivative under the abelian gauge transformation
 \be\label{gaugeSb}
 \delta b_{\alpha\dot\alpha} = \partial_\alpha\lambda_{\dot\alpha}\, , \qquad
 \delta \tilde A_{\dot\alpha\dot\beta} = 2\partial_{[\dot\alpha}\lambda_{\dot\beta]}\, ,
 \ee
which is an invariance of the field-strengths
 \be
 \tilde H_{\dot\alpha\dot\beta\gamma}  :=  \tilde \partial_{\dot\alpha} b_{\gamma\dot\beta} -
 \tilde\partial_{\dot\beta}b_{\gamma\dot\alpha} - \partial_\gamma\tilde A_{\dot\alpha\dot\beta}\, ,
 \qquad
 \tilde H_{\dot\alpha\dot\beta\dot\gamma} :=
  3 \tilde\partial_{\dot\alpha}\tilde A_{\dot\beta\dot\gamma}\, .
  \ee
The rescalings conspire such that $H_{\dot\alpha\dot\beta\dot\gamma}= \tilde H_{\dot\alpha\dot\beta\dot\gamma}$, from which it follows that
  \be\label{Hpi8}
 \frac{1}{3!} \varepsilon^{\dot\alpha\dot\beta\dot\gamma}
\tilde H_{\dot\alpha\dot\beta\dot\gamma} = - T^{-\nu} \pi_8\, .
\ee
This tells us,  firstly, that  we may trade the gauge-invariant part
of $\tilde A_{\dot\alpha\dot\beta}$ for $\pi_8$ and, secondly, that
this trade involves a factor of $T^{-\nu}$; in other words, there is
a $3$-vector field $\Lambda$ such that \be \tilde
A_{\dot\alpha\dot\beta} = 2\partial_{[\dot\alpha}
\Lambda_{\dot\beta]} + {\cal O}\left(T^{-\nu}\right)\, . \ee In
principle,  the ${\cal O}\left(T^{-\nu}\right)$ term can be
expressed in terms of $\pi_8$, but this expression would be
non-local on $M_3$. In addition, it is unclear to us how it could be
part of some $SO(8)$ invariant term. For this reason, among others
that we will encounter later, we shall be considering a $T\to\infty$
limit, and in this spirit we write \be \tilde
H_{\dot\alpha\dot\beta\gamma} = \tilde\partial_{\dot\alpha}
\left(b_{\gamma\dot\beta} - \partial_\gamma\Lambda_{\dot\beta}
\right) -
\tilde\partial_{\dot\beta}\left(b_{\gamma\dot\alpha}-\partial_\gamma\Lambda_{\dot\alpha}\right)
+  {\cal O}\left(T^{-\nu}\right)\, . \ee We see that $\Lambda$ is a
St\"uckelberg field: invariance of $\tilde
H_{\dot\alpha\dot\beta\gamma}$ under the abelian gauge
transformation   $\delta b_{\alpha\dot\alpha} = \partial_{\alpha}
\lambda_{\dot \alpha}$ is ensured by the  `shift'
$\delta\Lambda_{\dot\alpha} = \lambda_{\dot\alpha}$.  This gauge
invariance is therefore `spontaneously' broken by the St\"uckelberg
mechanism; we may choose to set $\Lambda=0$, thereby `fixing' this
gauge freedom\footnote{Alternatively, one can just write all the formulae to follow in terms of
 $\tilde{b}_{\alpha\dot\beta} := b_{\alpha\dot\beta} - \partial_\alpha \Lambda_{\dot\beta}$.}.
Notice, however, that there is
still an unbroken abelian  gauge invariance of $b$ viewed as a
2-vector-valued gauge potential on $M_3$.

With the rescalings as given, we also have
\begin{eqnarray}\label{Hdadbdg=}
H_{\dot{\alpha}\dot{\beta}{\gamma}} &=&   T^{- 2\nu/3}\tilde H_{\dot\alpha\dot\beta\gamma}
\, , \nonumber\\
\frac{1}{2}\varepsilon^{\beta\gamma}H_{\dot\alpha\beta\gamma} &=&  - T^{- \nu/3}\left[ \tilde\partial_{\dot\alpha} \phi^8 + 2T^{-\nu} \varepsilon^{\beta\gamma} \partial_\beta b_{\gamma\dot\alpha}\right] \, ,
\end{eqnarray}
and hence
 \be \label{Kda=} K_{\dot \alpha} = T^{1-{4\nu\over 3}}
\left[\tilde{K}_{\dot \alpha}+ {\cal O}(T^{-\nu})\right] \, , \qquad K_{\alpha} =
T^{1-\nu}\; \left[\tilde{K}_{\alpha}+ {\cal O}(T^{-\nu})\right] \, ,
 \ee
where, for the gauge choice $\Lambda=0$,
\be \bar e\label{tKda=}
\tilde{K}_{\dot \alpha} =  \tilde{\partial}_{\dot
\alpha}\phi^{\hat{I}}\, \pi_{\hat{I}} + \tilde{\partial}_{\dot
\alpha}\phi^{8}\, \pi_{8} - {1\over
2}\,\varepsilon^{{\alpha}{\beta}}
\varepsilon^{\dot{\beta}\dot{\gamma}\dot{\delta}}
(\tilde{\partial}{}_{\dot{\alpha}} b_{\alpha
\dot{\beta}}-\tilde{\partial}{}_{\dot{\beta}}  b_{\alpha
\dot{\alpha}})\, \tilde{\partial}{}_{\dot{\gamma}}
b_{\beta\dot{\delta}}+ {i\over 2}\bar{e} \, \tilde{\partial}_{\dot
\alpha}{\Psi}^T\Psi\, ,
 \ee
and
 \be \bar e
\label{tKa=} \tilde{K}_{\alpha}=  \tilde{P}_\alpha + \,
\varepsilon^{\dot{\alpha}\dot{\beta}\dot{\gamma}} \,
\tilde{\partial}_{\dot \alpha}{\Phi}^8\;
\tilde{\partial}_{\dot{\beta}}\, \, b_{\alpha\dot{\gamma}}\, ,
\qquad \tilde P_\alpha : = T^{\nu -1 } P_\alpha\, .
 \ee
Notice the absence of fermionic terms in $\tilde K_{\alpha}$; for finite  $T$ they  appear
with a $\partial_\alpha$  derivative but are suppressed in the $T\to\infty$ limit. As we
discuss below, the disappearance of $\bR^2$ derivatives is a general effect of the
limit we consider.

We are now in a position to see how an SDiff$_3$ gauge group will emerge in the
$T\to\infty$ limit. First, we define rescaled shift functions by
 \begin{eqnarray}\label{tsa=}
{s}{}^{\dot \alpha}=  T^{-2\nu/3} \tilde{s}{}^{\dot \alpha}\; , \qquad {s}{}^{ \alpha}=
T^{-\nu} \tilde{s}{}^{\alpha}\, .
\end{eqnarray}
As a consequence, we have
 \be
s^iK_i =  T^{1-2\nu}\left[\tilde{s}^{\dot
\alpha}\tilde{K}_{\dot \alpha} + \tilde{s}^{ \alpha} \tilde{K}_{ \alpha}
+ {\cal O}(T^{-\nu})\right]\, ,
 \ee
and
 \be\label{SDiff3}
\partial_i \left(\bar e s^i\right) = \tilde\partial_{\dot\alpha}\left(\bar e \tilde s^{\dot\alpha}\right)+
{\cal O}\left(T^{-\nu}\right)\, .
 \ee
In the $T\to\infty$ limit, the rescaled shift-function components $\tilde s^{\dot\alpha}$ satisfy a divergence-free condition, and the components $\tilde s^\alpha$ become {\it unconstrained} Lagrange multipliers that impose the constraint $\tilde K_\alpha=0$, which is trivially solved for $\tilde P_\alpha$. Thus
 \be\label{tKa=0}
\tilde{P}_\alpha = - \, \varepsilon^{\dot{\alpha}\dot{\beta}\dot{\gamma}} \,
\tilde{\partial}_{\dot \alpha}{\phi}^8\; \tilde{\partial}{}_{\dot{\beta}}\, \,
b_{\alpha\dot{\gamma}} + {\cal O}(T^{-\nu})\, .
\ee

Now we turn to the terms in the Hamiltonian. One such term is
\be\label{Psquared}
\frac{P_IP_I} {T\bar e} = T^{1-2\nu}
\left[ \bar e^{-1} \pi_{\hat I} \pi_{\hat I} +  \bar e^{-1} \tilde P_\alpha \tilde P_\alpha\right]\, .
\ee
Substitution for $\tilde P_\alpha$ yields a term that is not $SO(8)$ invariant but we still have
many other terms to consider.  For example,
 \begin{eqnarray}\label{T|5g|=}
\bar e^{-1}  T |{}^5g| &=& T^{1-2\nu}\left[\bar e^{-1} \det \left(\tilde{\partial}_{\dot{\alpha}}
\phi^{\hat{I}}\tilde{\partial}_{\dot{\beta}}\phi^{\hat{I}} \right) + {\cal O}\left(T^{-2\nu}\right)\right] \nn
&=& T^{1-2\nu} \left[ {1\over 3!} \bar e \left\{\phi^{\hat I}, \phi^{\hat J} , \phi^{\hat K}\right\}^2
+ {\cal O}\left(T^{-2\nu}\right)\right] \, ,
\end{eqnarray}
where we have defined the Nambu bracket  of functions $(F,G,H)$ by
\be
\left\{F,G,H\right\} = \bar e^{-1} \varepsilon^{\dot\alpha\dot\beta\dot\gamma} \tilde\partial_{\dot\alpha} F \, \tilde\partial_{\dot\beta} G \, \tilde\partial_{\dot\gamma} H\, .
\ee
Omitting the overall power of $T$, which is the same as in (\ref{dotAdA=}), we see that
 the leading term in the $T\to\infty$ limit is an $SO(7)$-invariant  potential  that can be expressed
 as  a sum of squares of  Nambu brackets  for the 7 scalar fields $\phi^{\hat I}$.

 A similar computation yields
 \begin{eqnarray}\label{gHH=}
{T|{}^5g| \over 3! \bar e}H_{ijk}H^{ijk} &=& T^{1- 2\nu }\left[ \bar e^{-1} \pi_8^2 +
 \bar e^{-1} \left(\varepsilon^{\dot{\alpha}\dot{\beta}\dot{\gamma}}\tilde{\partial}_{\dot{\alpha}}\phi^{\hat{I}}
\tilde{\partial}_{\dot{\beta}} \tilde{b}_{\gamma\dot{\gamma}}
\right)^2 \right. \nn
&& \left. + \ {1\over 2}\bar e \left\{\phi^8, \phi^{\hat I} , \phi^{\hat J}\right\}^2
+\  {\cal O}(T^{- \nu})\right]\, .
\end{eqnarray}
The first term on the right hand side provides the $SO(8)$ completion of the  $\pi^2$ term in  (\ref{Psquared}), and the second term provides the $SO(8)$ completion of  the $\tilde P_\alpha^2$ term in
(\ref{Psquared}).  The third term  provides the $SO(8)$ completion of the leading, potential, term
of (\ref{T|5g|=}).

We have now found all terms of leading order in an expansion in inverse powers of $T$ that survive
the truncation in which all fermion terms are omitted. The  $SO(8)$ invariance of this bosonic truncation
can be made manifest by defining an $SO(8)$-vector valued scalar field $\Phi$, and its conjugate momentum $\Pi$, by
 \be
 \Phi^{\tilde I} = \left(\phi^{\hat
I},\phi^8\right)\, , \qquad \Pi_{\tilde I} = \left(\pi_{\hat I},\pi_8\right)\, .
 \ee
 We shall postpone a presentation of the manifestly $SO(8)$ invariant results in this notation until we have dealt with the fermion terms.

\subsection{Fermions}

We have already seen in (\ref{kineticXP}) that there is a fermion bilinear `kinetic'
term, and in (\ref{tKda=}) that there is a fermion bilinear contribution to the
SDiff$_3$ constraint function. These fermion terms  are manifestly $SO(8)$, in fact
$SO(9)$, invariant. However, we still have to consider the fermion bilinear $T{\cal
C}_{++}$. In our rescaled variables this becomes
\begin{eqnarray}\label{finalferm}
T{\cal C}_{++} &=& T^{1-2\nu} \left[ i\varepsilon^{\dot{\alpha}\dot{\beta}\dot{\gamma}}
 \tilde{\partial}{}_{\dot{\alpha}}\, \,
b_{\alpha\dot{\beta}}\, \varepsilon^{{\alpha}{\beta}} \tilde{\partial}_{\dot \gamma} \Psi^T
\gamma^{\hat{\beta}}\Psi  + {i\over 2}
\varepsilon^{\dot{\alpha}\dot{\beta}\dot{\gamma}} \tilde{\partial}{}_{\dot{\alpha}}
{\phi}^{\hat{I}}\;\tilde{\partial}{}_{\dot{\beta}} {\phi}^{\hat{J}}\;
\tilde{\partial}_{\dot \gamma} \Psi^T \gamma_* \gamma^{\hat{I}\hat{J}}\Psi \right. \nonumber \\
&&\left.+ \ i \varepsilon^{\dot{\alpha}\dot{\beta}\dot{\gamma}}
\tilde{\partial}{}_{\dot{\alpha}} {\phi}^{8}\;\tilde{\partial}{}_{\dot{\beta}}
{\phi}^{\hat{J}}\; \tilde{\partial}_{\dot \gamma} \Psi^T \gamma^{\hat{J}}\Psi\; +{\cal
O}(T^{-\nu})\; \right] \, ,
\end{eqnarray}
where
\be
\gamma_*= {1\over 2}\varepsilon^{\alpha\beta}\gamma_{\alpha\beta}\, .
\ee
This is not obviously $SO(8)$ invariant. To show that it {\it is} $SO(8)$-invariant,
we must first decompose the $SO(9)$ spinor $\Psi$ into its irreducible $SO(8)$
representations. To do this we choose the
$SO(9)$ gamma matrices $\gamma^I= \left(\gamma^{\tilde I}, \gamma^9\right)$  to be
\be \label{Spin9=Spin8} \gamma^{\tilde{I}}= \left(\matrix{ 0 &
\rho^{\tilde{I}}_{A\dot{B}} \cr \tilde{\rho}{}^{\tilde{I}}_{\dot{A}B}  & 0}\right)\,
,\qquad  \gamma^9 =
  \left(\matrix{ \delta_{A{B}} & 0 \cr 0 & -\delta_{\dot{A}\dot B}
 }\right)\; ,
\ee where $\rho^{\tilde I}$  are $SO(8)$ `sigma'  matrices, with
transpose $\tilde\rho^{\tilde I}$; i.e. $\tilde{\rho}{}^{\tilde{I}}_{\dot{A}B} :=
{\rho}{}^{\tilde{I}}_{B\dot{A}}$.  These  $8\times 8$
matrices satisfy\footnote{It is  useful to keep in mind the $SO(7)$ invariant representation for the $SO(8)$ sigma
matrices in terms of octonionic structure constants, for which $\rho^8_{A\dot B} = \delta_{A\dot B}$ (see e.g. \cite{mM2-f}). }
 \be
\rho^{\tilde{I}}\tilde{\rho}^{\tilde{J}} + \rho^{\tilde{J}}\tilde{\rho}^{\tilde{I}}=
2\delta^{\tilde{I}\tilde{J}}\, 1_s\, , \qquad
\tilde\rho^{\tilde{I}}{\rho}^{\tilde{J}} + \tilde\rho^{\tilde{J}}{\rho}^{\tilde{I}}=
2\delta^{\tilde{I}\tilde{J}}\, 1_c\, ,
 \ee
where $1_s$ and $1_c$ are the identity matrices acting on ${\bf
8}_s$ and ${\bf 8}_c$ spinors. The $SO(8)$ generators acting on
${\bf 8}_s$ and ${\bf 8}_c$ spinors are
 \be
\rho^{\tilde I\tilde J}_{AB}\  := \left(\rho^{[\tilde
I}\tilde\rho^{\tilde J]}\right)_{AB}\, , \qquad \tilde\rho^{\tilde
I\tilde J}_{\dot A\dot B}  :=
 \left(\tilde\rho^{[\tilde I}\rho^{\tilde J]}\right)_{\dot A\dot B}\, .
 \ee
In this basis, an $SO(9)$ spinor $\Psi$ takes the form
 \be
 \Psi =
\pmatrix{\chi_A\cr \tilde\chi_{\dot A}}\, ,
 \ee
where $\chi$ and $\tilde\chi$ are $SO(8)$ spinors in, respectively,  the ${\bf 8}_s$ and ${\bf 8}_c$ representations. We may trade these spinors for a doublet of  ${\bf 8}_s$  spinors, which we may view
as an ${\bf 8}_s$-plet of real  $SO(1,2)$ spinors,
\be
\psi_A = \pmatrix{\chi_A \cr \left(\rho^8\tilde \chi\right)_A}\, \qquad (A=1,\dots,8).
\ee

To facilitate this new interpretation, we introduce {\it irreducible} $2\times 2$ Dirac matrices $\tilde\gamma^\mu$
satisfying
\be
\left[\tilde\gamma^\mu, \tilde\gamma^\nu \right]_+ = \eta^{\mu\nu}\, , \qquad (\mu=0,1,2).
\ee
For a `mostly minus' signature convention (in accord with the choice made for the
11-dimensional Minkowski metric) these matrices are pure imaginary in a Majorana basis. A
convenient choice is
 \be\label{gamma1+2}
 \tilde\gamma^0 = \tau_2\, , \qquad
\tilde\gamma^1 = i\tau_1\, , \qquad \tilde\gamma^2 = i\tau_3 \; . \ee where
$(\tau_1,\tau_2,\tau_3)$ are the Hermitian Pauli matrices. The $2\times 2$ charge
conjugation matrix $c$ can be chosen to be $\tilde\gamma^0$, in which case
$c\tilde\gamma^0=1_2$ and both $c\tilde\gamma^1 = \tau_3$ and $c\tilde\gamma^2 =
-\tau_1$ are real symmetric matrices. For this choice, the Dirac conjugate of an
$SO(1,2)$ Majorana spinor $\psi$ is \be \bar\psi = \psi^t \tilde\gamma^0\, , \ee where
the superfix $t$ indicates the transpose of the 2-component spinor.  As $\rho^8$
squares to the identity, the `kinetic'  term for $\Psi$ can now be written as
\be\label{Fermi2=SO(8)} \frac{i}{2} \dot\Psi^T \Psi = -\frac{i}{2} \bar{\psi}_A
\tilde\gamma^0\dot\psi_A \, . \ee In this new notation, (\ref{finalferm}) becomes
 \be
T{\cal C}_{++} = T^{1-2\nu} i\varepsilon^{\dot{\alpha}\dot{\beta}\dot{\gamma}}\left[
\tilde{\partial}{}_{\dot{\alpha}} b_{\alpha\dot{\beta}}\, \left( \tilde{\partial}_{\dot
\gamma} \bar{\psi}{}_A \check{\gamma}^{{\alpha}}\psi_A\right) - {i\over 2}
 \tilde{\partial}{}_{\dot{\alpha}}
{\phi}^{\tilde{I}}\;\tilde{\partial}{}_{\dot{\beta}} {\phi}^{\tilde{J}}\;
\tilde{\partial}_{\dot \gamma} \bar{\psi}{}_A \rho^{\tilde{I}\tilde{J}}_{AB}\psi_B \;
+{\cal O}(T^{-\nu})\; \right]\, .
 \ee
Finally, we can now rewrite the expression (\ref{tKda=})  for $\tilde K_{\dot\alpha}$
in manifestly $SO(8)$ invariant form as
 \be \label{Ktil=}
\bar e \tilde{K}_{\dot\alpha} = \tilde\partial_{\dot\alpha}
\Phi^{\tilde I}\Pi_{\tilde I} - {1\over
2}\varepsilon^{{\alpha}{\beta}}  \,
\varepsilon^{\dot{\beta}\dot{\gamma}\dot{\delta}}
(\tilde{\partial}{}_{\dot{\alpha}} b_{\alpha
\dot{\beta}}-\tilde{\partial}{}_{\dot{\beta}}  b_{\alpha
\dot{\alpha}})\, \tilde{\partial}{}_{\dot{\gamma}}
b_{\beta\dot{\delta}} -  {i\over 2}\bar{e} \,
\bar{\psi}_A\tilde\gamma^0\tilde{\partial}_{\dot \alpha} \psi_A\, ,
\ee

\section{Carrollian BLG}
\setcounter{equation}{0}

We have now shown that  the Lagrangian density of the light-cone
gauge fixed M5-brane  can be written, after some further gauge
fixing, in the form \be {\cal L}_{M5} = T^{1-2\nu} \left[
\tilde{\cal L} + {\cal O}\left(T^{-\nu}\right)\right] \ee where
$\tilde{\cal L}$  is an $SO(8)$ invariant  constructed from rescaled
fields that are functions of coordinates $x^\mu=(t,x^\alpha)$  of a
3-dimensional Minkowski spacetime, and  rescaled coordinates
$\tilde\sigma^{\dot\alpha}$  for a compact 3-space. As we observed
previously, the overall factor of $T^{1-2\nu}$ cancels from the
action $I_{M5}$ when $\nu=1/4$, and in this case we can define
 \be
\tilde I :=  \lim\limits_{T\to\infty} I_{M5} =  \int d^2x \left[
\oint d^3\tilde\sigma \tilde{\cal L}\right]
 \ee
Putting together the results of the previous section, we see that
 \be \label{LBLG'=}
 \tilde{\cal L} = \dot{\Phi}^{\tilde{I}} \Pi_{\tilde{I}} -
 {i\over 2}\bar{e}\, \bar{\psi}_A\tilde{\gamma}^0 \dot {\psi}_A
  - {1\over 2}\varepsilon^{{\alpha}{\beta}}\varepsilon^{\dot{\alpha}\dot{\beta}\dot{\gamma}}
  \dot{b}{}_{{\alpha}\dot{\alpha}}
\tilde{\partial}{}_{\dot{\beta}} b_{\beta\dot{\gamma}}+ \bar{e}\, \tilde s^{\dot\alpha}
\tilde K_{\dot\alpha} - \tilde {\cal H} \ee where
\begin{eqnarray} \label{F2=SO(8)NB}
\tilde{\cal H} &=& \bar e^{-1} \left[
\Pi_{\tilde{I}} \Pi_{\tilde{I}} +
\left(\varepsilon^{\dot\alpha\dot\beta\dot\gamma}\tilde\partial_{\dot\alpha}\Phi^{\tilde
I} \tilde \partial_{\dot\beta}b_{\beta\dot\gamma}\right)^2 \right] + {1\over
3!} \bar e \sum\limits_{\tilde{I},\tilde{J},\tilde{K}}\left\{{\Phi}^{\tilde{I}}\,
, \, {\Phi}^{\tilde{J}}\, , \, {\Phi}^{\tilde{K}}\right\} ^2 \nonumber \\
&& + \ i\varepsilon^{\dot\alpha\dot\beta\dot\gamma}
\tilde\partial_{\dot\alpha}b_{\alpha\dot\beta} \left(\tilde\partial_{\dot\gamma}
\bar\psi_A \tilde\gamma^\alpha \psi_A\right) - {1\over 2}\bar{e} \left\{
{\Phi}^{\tilde{I}}\, , \,  {\Phi}^{\tilde{J}}\, , \, \bar{\psi}{}_A\right\} \,
\rho^{\tilde{I}\tilde{J}}_{AB}\psi_B\, .
\end{eqnarray}
The constraint function $\tilde K_{\dot\alpha}$ is given by (\ref{Ktil=}) and the
Lagrange multiplier for this constraint  satisfies \be\label{scon}
\tilde\partial_{\dot\alpha}\left(\bar e \tilde s^{\dot\alpha}\right) =0\, , \ee which
we may solve, locally, by  writing \be\label{timeb} \bar e \tilde s^{\dot\alpha} =
-2B_t^{\dot\alpha}\, , \qquad B_t^{\dot\alpha} :=
\varepsilon^{\dot\alpha\dot\beta\dot\gamma} \tilde\partial_{\dot\beta} b_{t\dot\gamma}
\, . \ee As we shall see shortly, the unconstrained Lagrange multiplier
$b_{t\dot\alpha}$ , which is  only defined up
to a gauge transformation with $\delta b_{t\dot\alpha}= \dot\lambda_{\dot\alpha}$, may
be combined with $b_{\alpha\dot\alpha}$ to form a 3-vector valued $SO(1,2)$-vector
$b_{\mu\dot\alpha}$, and similarly for $B_t^{\dot\alpha}$, which is the time component
of a 3-vector valued $SO(1,2)$-vector $B_{\mu}^{\dot\alpha}$.

We now rewrite the Lagrangian density as
 \be \tilde{\cal L} =
D_t\Phi^{\tilde I} \Pi_{\tilde I} -\frac{i}{2}\bar e\, \bar\psi_A
\tilde\gamma^0D_t\psi_A - \tilde{\cal H}  + \tilde {\cal L}_{CS}\, ,
 \ee
where
 \be \tilde{\cal L}_{CS} =  - \frac{1}{2}
\varepsilon^{\alpha\beta}\varepsilon^{\dot\beta\dot\gamma\dot\delta} \left[ \dot
b_{\alpha\dot\beta} -{2\over
\bar{e}}\varepsilon^{\dot\alpha\dot\lambda\dot\eta}\tilde\partial_{\dot\lambda}b_{t
\dot\eta}
\left(\tilde\partial_{\dot\alpha}b_{\alpha\dot\beta}-\tilde\partial_{\dot\beta}b_{\alpha\dot\alpha}\right)\right]
\tilde\partial_{\dot\gamma}b_{\beta\dot\delta} \, . \ee
The `CS' subscript  will be
explained in the subsection to follow.  The SDiff$_3$ covariant time derivative  $D_t$
is now defined, locally, on any of the fields $(\Phi^{\tilde I},\psi_A)$, which we
denote collectively by $\Xi$, as \be\label{covdiv0} D_t \Xi  := \partial_t \Xi + \tilde
s^{\dot\alpha}\partial_{\dot\alpha}\Xi  = \partial_t \Xi -
 2\bar{e}^{-1} \varepsilon^{\dot\alpha\dot\beta\dot\gamma}\tilde\partial_{\dot\alpha}\Xi\,
\tilde \partial_{\dot\beta}b_{t\dot\gamma}\, .
\ee

Conspicuous by their absence are any $\bR^2$ derivatives of  any of the fields, scalar,
spinor or gauge. In the Nambu bracket realization of the BLG theory \cite{BL07}, these
derivatives occur together with minimal coupling terms, such that both are taken into
account via the `covariant derivative' \cite{BLGM5}
 \be
 D_\alpha\,  \Xi = \partial_\alpha \Xi -  2\bar{e}^{-1} \varepsilon^{\dot\alpha\dot\beta\dot\gamma}\tilde\partial_{\dot\alpha} \, \Xi \,
\tilde \partial_{\dot\beta}b_{\alpha\dot\gamma} =
\partial_\alpha \Xi - 2 \left\{\Xi,  b_{\alpha\dot\beta},
\tilde\sigma^{\dot\beta}\right\}\,.
 \ee
The `connection' terms in this derivative yield the terms in  (\ref{LBLG'=}) that couple $b$ to $\Phi$ and $\psi$.  As can be seen by comparison with (\ref{covdiv0}), the covariant derivatives $(D_t,D_\alpha)$  are the components of the 3-vector covariant derivative
\be
D_\mu \Xi = \partial_\mu \Xi -2 \left\{ \Xi , b_{\mu\dot\alpha}, \tilde\sigma^{\dot\alpha}\right\} \, , \qquad
b_{\mu\dot\alpha}=(b_{t\dot\alpha}, b_{\alpha\dot\alpha})\, .
\ee

If we now eliminate the 8-momentum, we arrive at the Lagrangian density
\begin{eqnarray}\label{LBLG=}
 \tilde{\cal L}&=& {\bar e\over 4}\left[ \left|D_t\Phi\right|^2 -
 \left|\left(D_\alpha - \partial_\alpha\right)\Phi\right|^2\right]
 + \tilde {\cal L}_{CS} \nn
&&- \ {i\over 2}\bar e \left[ \bar\psi_A \tilde\gamma^0 D_t \psi_A + \bar\psi_A
 \tilde\gamma^\alpha \left(D_\alpha -\partial_\alpha\right) \psi_A\right]  \\
&& - {\bar  e \over 3!}
 \sum_{\tilde{I},\tilde{J},\tilde{K}}\left\{{\phi}^{\tilde{I}}\,
, \, {\phi}^{\tilde{J}}\, , \, {\phi}^{\tilde{K}}\right\} ^2 + \ {1\over 2}\bar{e}\,
\left\{ {\Phi}^{\tilde{I}}\, , \, {\Phi}^{\tilde{J}}\, , \, \bar{\psi}{}_A\right\} \,
\rho^{\tilde{I}\tilde{J}}_{AB}\psi_B \, ,  \nonumber
\end{eqnarray}
where we now recognize $D_\mu=(D_t,D_\alpha)$ as an SDiff$_3$ covariant derivative (we elaborate on the SDiff$_3$ gauge invariance in the subsection to follow).
Leaving aside the ${\cal L}_{CS}$ term, it is now clear that we have found a Lagrangian density for a
3-dimensional gauge theory in which the $SO(1,2)$ Lorentz invariance is
 broken by the subtraction of all terms with space derivatives. This is also true of the `CS'
 term, as may be seen by considering the manifestly $SO(1,2)$-invariant Lagrangian
 density \cite{BLGM5-flux}
 \be {\cal L}_{CS} = -\frac{1}{2}\varepsilon^{\mu\nu\rho}
\left[\partial_\mu b_{\nu\dot\gamma} + {2\over 3\bar{e}}\,
\varepsilon_{\dot\alpha\dot\beta\dot\gamma}
B_\mu^{\dot\alpha}B_\nu^{\dot\beta}\right] B_\rho^{\dot\gamma}\, ,
 \ee
where, by definition,
 \be\label{Bda:=}
B_\mu ^{\dot\alpha} := \varepsilon^{\dot\alpha\dot\beta\dot\gamma}
\tilde\partial_{\dot\beta}b_{\mu\dot\gamma}\, ,  \qquad
\varepsilon^{\dot\alpha\dot\beta\dot\gamma}\varepsilon_{\dot\delta\dot\epsilon\dot\eta}
= 3! \delta^{[\dot\alpha}_{\dot\delta}
\delta^{\dot\beta}_{\dot\epsilon}\delta^{\dot\gamma]}_{\dot\eta}\, .
 \ee
It is straightforward to show, ignoring total space derivatives, that \be
{\cal L}_{CS} = \tilde {\cal L}_{CS} - b_{t\dot\alpha} \,\varepsilon^{\alpha\beta}
\partial_\alpha B^{\dot\alpha}_\beta\, . \ee

We have now shown that the dynamics of an  M5-brane of topology $\bR^2\times M_3$  is
governed, in the particular hypertensile limit that we have taken,  by a
$(1+2)$-dimensional field theory. The fields of this theory, which are tensor-valued in
an `auxiliary' closed 3-manifold $M_3$, and multiplets of $SO(8)$, consist of an ${\bf
8}_v$-plet  of scalar fields ($\Phi$), an ${\bf 8}_s$-plet of two-component real
$Sl(2;\bR)$ spinor fields ($\psi$), both scalars on $M_3$, and an $SO(8)$-singlet
vector field $b$ that is also a vector potential on $M_3$. This field theory is
precisely the BLG theory in its Nambu bracket  realization except that all space
derivatves are absent. The relativistic, SO(1,2) invariance is broken down to SO(2) by
this absence of spatial derivatives. It should {\it not}  be thought that we
have found a dimensional reduction of the BLG theory to one dimension (time) because
{\it the fields were never assumed to be independent of the $\bR^2$ coordinates}.
Instead, what we have found is the  Carrollian limit of  the Nambu-bracket BLG theory,
in which the speed of light has been taken to zero; this has precisely the effect of
suppressing all spatial derivatives.

\subsection{SDiff$_3$ gauge invariance}

The BLG theory is an SDiff$_3$  gauge theory because it is invariant under the  gauge
transformations\footnote{We recall that $\Xi$ stands collectively for the fields $(\Phi^{\tilde I},\psi_A)$.}
\be\label{SDiffgauge}
 \delta\Xi = - \tilde\zeta^{\dot\alpha}\tilde\partial_{\dot\alpha}\, \Xi , \qquad
 \delta b_{\dot \alpha} = d\omega_{\dot\alpha} - \tilde\zeta^{\dot\beta}\tilde \partial_{\dot\beta}b_{\dot\alpha} - \tilde\partial_{\dot\alpha}\tilde\zeta^{\dot\beta}\, b_{\dot\beta}\, ,
\ee where \be \tilde \zeta^{\dot\alpha} = -2 \bar e^{-1}
\varepsilon^{\dot\alpha\dot\beta\dot\gamma}
\tilde\partial_{\dot\beta}\omega_{\dot\gamma}\, . 
\ee 
This defines
$\omega_{\dot\alpha}$ in terms of $\tilde\zeta^{\dot\alpha}$ only up to the addition of
$\tilde\partial_{\dot\alpha}\varsigma$, for any  scalar $\varsigma$, but this addition leads to
an SDiff$_3$ transformation of $b_{\dot\alpha}$ that is equivalent to the one given
once account is taken of the unbroken abelian gauge invariance of $b_{\dot\alpha}$.
Note that
 \be \tilde\partial_{\dot\alpha}\left( \bar e \tilde{\zeta}{}^{\dot\alpha}\right) \equiv 0\, . 
 \ee
Let $B^{\dot\alpha} = dx^\mu B_\mu^{\dot\alpha}$, where $B_\mu^{\dot\alpha}$ is as 
defined in (\ref{Bda:=}); then 
 \be 
 \delta \left(-2\bar e^{-1} B^{\dot\alpha}\right) = d\tilde\zeta^{\dot\alpha} + \left(-2\bar e^{-1}B^{\dot\beta}\right)\tilde\partial_{\dot\beta}\tilde\zeta^{\dot\alpha} -\tilde\zeta^{\dot\beta}\tilde\partial_{\dot\beta}\left(-2\bar e^{-1}B^{\dot\alpha}\right)\, . 
 \ee
Recalling (\ref{timeb}), we see that this includes the transformation
 \be
\tilde s^{\dot\alpha} = d\tilde\zeta^{\dot\alpha} + \left[\tilde s,
\tilde\zeta\right]^{\dot\alpha}\, ,
\ee
where $[,]$ is the Lie bracket of  vector fields on $M_3$.  This transformation follows from the SDIff$_5$ transformation of $s^i$ given in (\ref{infinitesimal}) after taking the hypertensile limit described in section \ref{sec:hyper} with
\be
\zeta^{\dot\alpha}= T^{-\frac{2\nu}{3}} \tilde\zeta^{\dot\alpha}\, .
\ee
The SDiff$_3$ transformation of  the space components $b_{\alpha\dot\alpha}$ of
$b_{\dot\alpha}$ may be similarly deduced from  those of $A$ given in (\ref{infinitesimal}),  and
the result (in the gauge $\Lambda_{\dot\alpha}=0$, agrees with  (\ref{SDiffgauge})  if one drops the spatial derivative term $\partial_\alpha \omega_{\dot\alpha}$. 

With the exception of the ${\cal L}_{CS}$ term, the SDiff$_3$ gauge invariance of  all
terms of the BLG action is manifest (because it is constructed using the covariant
derivative $D$). The ${\cal L}_{CS}$ term is a type of Chern-Simons term in the sense
that its variation is a total  spacetime derivative\footnote{We may ignore total $M_3$ derivatives
since $M_3$ has no boundary.}. Alternatively, one may verify  the
covariance of the functional derivative with respect to $b_{\dot\alpha}$ of the  `CS'
action functional obtained by integration of ${\cal L}_{CS}$. This functional
derivative is proportional to the field-strength two-form\footnote{ Here, as in
(\ref{R4:=dC=}), we use the differential form conventions with exterior derivative
acting `from the right'. }
 \be\label{F:=}
 F^{\dot\alpha} := dB^{\dot \alpha} + 2B^{\dot\beta}\tilde\partial_{\dot\beta}
\left(\bar e^{-1} B^{\dot\alpha}\right)\, .
 \ee
Note that this field-strength 2-form is {\it not} of Yang-Mills type, because we are
dealing with an `exotic' gauge theory, and ${\cal L}_{CS}$ is therefore not a
Chern-Simons term in the usual sense of this term. However, it has the properties
required for SDiff$_3$ invariance because the transformation of $B$ induces the
transformation\footnote{Recall that $\bar{e}$ is a density on $M_3$ which is constant in
2--space and time, so that $d\bar{e}=0$.}
\be \delta F^{\dot\alpha} =
2\rho^{\dot\beta}\tilde\partial_{\dot\beta}\left(\bar e^{-1} F^{\dot\alpha}\right) -
2F^{\dot\beta}\tilde\partial_{\dot\beta}\left(\bar e^{-1}\rho^{\dot\alpha}\right) \, .
\ee
In other words, $F^{\dot\alpha}$ transforms covariantly under an  SDiff$_3$ gauge transformation-invariant, as claimed. In particular, the equation $F=0$ is SDiff$_3$ invariant.

\section{ Conclusions}
\setcounter{equation}{0}

We have presented the light-cone gauge fixed action for the M5-brane in the
11-dimensional Minkowski vacuum of M-theory. As expected from earlier results,  it has
an `exotic'  SDiff$_5$ gauge invariance. By  considering an M5-brane of topology
$\bR^2\times M_3$, for  some closed 3-manifold $M_3$, we found a $(1+2)$ dimensional
Minkowski space field theory, which is plausibly related to the recent `BLG' multiple
M2-brane model because an M5-brane may contain `dissolved' M2-branes. Crucially, the
BLG model has an $SO(8)$ invariance whereas only an $SO(7)$ invariance is guaranteed by
the M5 construction.  We found a limit, formally one of infinite M5 tension $T$
although the fields and coordinates were first scaled by powers of $T$, in which the
$SO(7)$ invariance is enhanced to $SO(8)$. In the same limit the partially gauge-fixed
SDiff$_5$ invariance is reduced to an SDiff$_3$ invariance and a BLG-like theory
emerges, complete with the expected potential term. However, the limit also suppresses
$\bR^2$ derivatives. Starting from the BLG theory, one can achieve the same supression
of spatial derivatives by taking a `Carrollian' limit, in which limit the speed of light
is zero.  We should point out that our  Carrollian limit of the (super)conformal BLG
theory is not itself conformal, although it is likely invariant under the contraction
of the (super)conformal group that is implied by the contraction of its Lorentz subgroup to
the Carroll group.

An interesting fact  (which we passed over previously for the sake of simplicity of
presentation) is that essentially the same results may be obtained by a  {\it zero
tension}  limit if one choose the parameter $\nu$ defining the various rescalings to be
negative. In this case, the fields and coordinates have `peculiar' dimensions and the
overall factor of $T^{1-2\nu}$ multiplying the leading term in the Lagrangian density
does not cancel in the action; one must rescale the action before taking the $T\to0$
limit (as is done to define Virasoro generators and BRST charge of a tensionless limit of the
string in \cite{Sagnotti+Mirian-Bonelli}). We do not know whether this fact is
of any significance but,  in light of it,  it is worth recalling that the Carroll group
arises naturally as the symmetry of a null brane in one higher dimension
\cite{Gibbons:2003gb}.

An obvious question is whether there is some other limit in which precisely the BLG
theory emerges. We cannot say for sure but we consider this unlikely for various
reasons.  To start with,  the symmetry algebra of the M5-brane in the Minkowski vacuum
of M-theory is an 11-dimensional super-Poincar\'e symmetry with tensor charges, and
neither this algebra nor any of its contractions  contains  the algebra of $OSp(8|4)$,
which is the symmetry supergroup of the BLG theory.  From this viewpoint, a better
starting point might be an M5-brane in the $adS_4\times S^7$ vacuum of
M-theory, because an M5-brane in this background is $OSp(8|4)$ invariant  \cite{KKKTP},
but it remains to be seen whether this will work. If it does, then it is likely that
the limit of  infinite adS radius,  in which the $adS_4\times S^7$ vacuum degenerates
to the 11-dimensional Minkowski  vacuum, will correspond to the Carrollian limit of the
`holographic'  BLG theory.

Another obvious question is whether analogous results might emerge by considering M5-branes
of other topologies, for example $S^1\times M_4$ for some closed 4-manifold $M_4$. One
might imagine that this could be related to some `exotic' $(1+1)$-dimensional
gauge theory based on a Filippov 4-algebra. However, all we were able to find was a
version of the $D4$-brane action in which all fields depend on a 5th space coordinate,
but without derivatives with respect to it. Another possibility is an M5-brane of
topology $\bR^3 \times M_2$; in this case there are many possibilities for rescaling
fields and therefore, potentially, there are many possible limits. We hope to
report on this case in a future publication.

\bigskip
\noindent
\section*{Acknowledgments}

The authors thank  Gary Gibbons for helpful discussions.  The work of  IAB  was
partially supported by research grants from the Spanish MEC (FIS2005-02761),
 the INTAS (2006-7928), the Ukrainian National Academy of Sciences and Russian  RFFI
grant 38/50--2008. PKT is supported by an ESPRC Senior Research Fellowship.


{\small
\begin{thebibliography}{99}

\bibitem{Filippov} V.T. Filippov, ``n-Lie algebras'', Sib. Mat. Zh., {\bf 26},
No 6, 126-140 (1985).

 \bibitem{Tak}
 L.~Takhtajan,
  ``On Foundation Of The Generalized Nambu Mechanics (Second Version),''
  Commun.\ Math.\ Phys.\  {\bf 160}, 295 (1994)
  [arXiv:hep-th/9301111]; \\
  %%CITATION = CMPHA,160,295;%%
 J.~A.~de Azc\'arraga, A.~M.~Perelomov and J.~C.~Perez Bueno,
 ``The Schouten-Nijenhuis
bracket, cohomology and generalized Poisson
  structures,''
  J.\ Phys.\ A  {\bf 29}, 7993-8010 (1996)
  [arXiv:hep-th/9605067];
  %%CITATION = JPAGB,A29,7993;%%
J.A. de Azc\'arraga, J.C. Perez Bueno, ``Higher order simple Lie algebras,''
Commun.Math.Phys.184:669-681,1997 [arXiv:hep-th/9605213];
%%CITATION = CMPHA,184,669;%%
J.A. de Azc\'arraga, J.M. Izquierdo, J.C. Perez Bueno, ``On the generalizations of
Poisson structures,'' J.Phys. A30, L607-L616 (1997). [arXiv: hep-th/9703019]
%%CITATION = JPAGB,A30,L607;%%


\bibitem{BL07}
  J.~Bagger and N.~Lambert,
``Modeling multiple M2's,''
  Phys.\ Rev.\  D {\bf 75}, 045020 (2007)
  [arXiv:hep-th/0611108];
  %%CITATION = PHRVA,D75,045020;%%
  ``Gauge Symmetry and Supersymmetry of Multiple M2-Branes,''
  Phys.\ Rev.\  D {\bf 77}, 065008 (2008)
  [arXiv:0711.0955 [hep-th]];
  %%CITATION = PHRVA,D77,065008;%%
``Comments On Multiple M2-branes,''
  JHEP {\bf 0802}, 105 (2008)
  [arXiv:0712.3738 [hep-th]].
  %%CITATION = JHEPA,0802,105;%%


\bibitem{G07}
A.~Gustavsson,
  ``Algebraic structures on parallel M2-branes,''
  arXiv:0709.1260 [hep-th];
  %%CITATION = ARXIV:0709.1260;%%
A.~Gustavsson,
  ``Selfdual strings and loop space Nahm equations,''
  JHEP {\bf 0804}, 083 (2008)
  [arXiv:0802.3456 [hep-th]].
  %%CITATION = JHEPA,0804,083;%%



%\cite{Schwarz+08}
\bibitem{Schwarz+08}
M.~A.~Bandres, A.~E.~Lipstein and J.~H.~Schwarz,
  ``N = 8 Superconformal Chern--Simons Theories,''
  JHEP {\bf 0805}, 025 (2008)
  [arXiv:0803.3242 [hep-th]].
  %%CITATION = JHEPA,0805,025;%%


\bibitem{Cherkis08}
S.~Cherkis and C.~Saemann,
  ``Multiple M2-branes and Generalized 3-Lie algebras,''
  arXiv:0807.0808 [hep-th].
  %%CITATION = ARXIV:0807.0808;%%


  %\cite{Lambert:2008et}
\bibitem{Lambert:2008et}
  N.~Lambert and D.~Tong,
  ``Membranes on an Orbifold,''
  arXiv:0804.1114 [hep-th].
  %%CITATION = ARXIV:0804.1114;%%

  %\cite{Distler:2008mk}
\bibitem{Distler:2008mk}
  J.~Distler, S.~Mukhi, C.~Papageorgakis and M.~Van Raamsdonk,
  ``M2-branes on M-folds,''
  JHEP {\bf 0805} (2008) 038
  [arXiv:0804.1256 [hep-th]].
  %%CITATION = JHEPA,0805,038;%%

   %\cite{Gauntlett:2008uf}
\bibitem{Gauntlett:2008uf}
  J.~P.~Gauntlett and J.~B.~Gutowski,
  ``Constraining Maximally Supersymmetric Membrane Actions,''
  arXiv:0804.3078 [hep-th].
  %%CITATION = ARXIV:0804.3078;%%


\bibitem{Papadopoulos08}
G.~Papadopoulos,
  ``M2-branes, 3-Lie Algebras and Plucker relations,''
  JHEP {\bf 0805}, 054 (2008)
  [arXiv:0804.2662 [hep-th]];
  %%CITATION = JHEPA,0805,054;%%
 %G.~Papadopoulos,
  ``On the structure of k-Lie algebras,''
  Class.\ Quant.\ Grav.\  {\bf 25}, 142002 (2008)
  [arXiv:0804.3567 [hep-th]].
  %%CITATION = CQGRD,25,142002;%%


%\cite{Russo08}
\bibitem{Russo08}
J.~Gomis, G.~Milanesi and J.~G.~Russo,
  ``Bagger-Lambert Theory for General Lie Algebras,''
 JHEP {\bf 0806}, 075 (2008)
  [arXiv:0805.1012 [hep-th]].
  %%CITATION = JHEPA,0806,075;%%
\\
S.~Benvenuti, D.~Rodriguez-Gomez, E.~Tonni and H.~Verlinde,
  ``N=8 superconformal gauge theories and M2 branes,''
  arXiv:0805.1087 [hep-th];
  %%CITATION = ARXIV:0805.1087;%%
%{\bf *****}
  \\ J.~Gomis, D.~Rodriguez-Gomez, M.~Van Raamsdonk and H.~Verlinde,
  ``Supersymmetric Yang-Mills Theory From Lorentzian Three-Algebras,''
  arXiv:0806.0738 [hep-th];
  %%CITATION = ARXIV:0806.0738;%%
%{\bf *****}
\\ H.~Verlinde,
  ``D2 or M2? A Note on Membrane Scattering,''
  arXiv:0807.2121 [hep-th].
  %%CITATION = ARXIV:0807.2121;%%


%\cite{BLGM5-M2D2}
\bibitem{BLGM5-M2D2}
  P.~M.~Ho, Y.~Imamura and Y.~Matsuo,
  ``M2 to D2 revisited,''
 JHEP {\bf 0807}, 003 (2008)
  [arXiv:0805.1202 [hep-th]].
  %%CITATION = JHEPA,0807,003;%%



%\cite{JHS-BLG2}
\bibitem{JHS-BLG2}
M.~A.~Bandres, A.~E.~Lipstein and J.~H.~Schwarz,
  ``Ghost-Free Superconformal Action for Multiple M2-Branes,''
  arXiv:0806.0054 [hep-th];
  %%CITATION = ARXIV:0806.0054;%%
  \\
 J.~Gomis, D.~Rodriguez-Gomez, M.~Van Raamsdonk and H.~Verlinde,
  ``The Superconformal Gauge Theory on M2-Branes,''
  arXiv:0806.0738 [hep-th].
  %%CITATION = ARXIV:0806.0738;%%

%\cite{D2toD2}
\bibitem{D2toD2}
B.~Ezhuthachan, S.~Mukhi and C.~Papageorgakis,
  ``D2 to D2,''
  JHEP {\bf 0807}, 041 (2008)
  [arXiv:0806.1639 [hep-th]].
  %%CITATION = JHEPA,0807,041;%%

     \bibitem{Eric+0608}
 E.~A.~Bergshoeff, M.~de Roo, O.~Hohm and D.~Roest,
  ``Multiple Membranes from Gauged Supergravity,''
  arXiv:0806.2584 [hep-th].
  %%CITATION = ARXIV:0806.2584;%%



\bibitem{BLG}
 M.~Van Raamsdonk,
  ``Comments on the Bagger-Lambert theory and multiple M2-branes,''
  JHEP {\bf 0805}, 105 (2008)
  [arXiv:0803.3803 [hep-th]].
  %%CITATION = JHEPA,0805,105;%%
\\
U.~Gran, B.~E.~W.~Nilsson and C.~Petersson,
  ``On relating multiple M2 and D2-branes,''
  arXiv:0804.1784 [hep-th].
  %%CITATION = ARXIV:0804.1784;%%
\\
 P.~M.~Ho, R.~C.~Hou and Y.~Matsuo,
  ``Lie 3-Algebra and Multiple M2-branes,''
  JHEP {\bf 0806}, 020 (2008)
  [arXiv:0804.2110 [hep-th]].
  %%CITATION = JHEPA,0806,020;%%
\\ E.~A.~Bergshoeff, M.~de Roo and O.~Hohm,
  ``Multiple M2-branes and the Embedding Tensor,''
   Class.\ Quant.\ Grav.\  {\bf 25}, 142001 (2008)
  [arXiv:0804.2201 [hep-th]].
  %%CITATION = CQGRD,25,142001;%%
  \\
   K.~Hosomichi, K.~M.~Lee and S.~Lee,
  ``Mass-Deformed Bagger-Lambert Theory and its BPS Objects,''
  arXiv:0804.2519 [hep-th].
  %%CITATION = ARXIV:0804.2519;%%
\\
  S.~Banerjee and A.~Sen,
  ``Interpreting the M2-brane Action,''
  arXiv:0805.3930 [hep-th];
  %%CITATION = ARXIV:0805.3930;%%
\\
J.~Figueroa-O'Farrill, P.~de Medeiros and E.~Mendez-Escobar,
  ``Lorentzian Lie 3-algebras and their Bagger-Lambert moduli space,''
  arXiv:0805.4363 [hep-th];
  %%CITATION = ARXIV:0805.4363;%%
%{\bf *****}
``Metric Lie 3-algebras in Bagger-Lambert theory,''
  arXiv:0806.3242 [hep-th].
  %%CITATION = ARXIV:0806.3242;%%

\bibitem{BLG-Morozov}
 A.~Morozov,
  ``On the Problem of Multiple M2 Branes,''
  JHEP {\bf 0805}, 076 (2008)
  [arXiv:0804.0913 [hep-th]].
  %%CITATION = JHEPA,0805,076;%%
\\
A.~Morozov,
  ``From Simplified BLG Action to the First-Quantized M-Theory,''
  arXiv:0805.1703 [hep-th],
  %%CITATION = ARXIV:0805.1703;%%
and refs. therein.



  %\cite{Krishnan:2008zm}
\bibitem{Krishnan:2008zm}
  C.~Krishnan and C.~Maccaferri,
  ``Membranes on Calibrations,''
  JHEP {\bf 0807}, 005 (2008)
  [arXiv:0805.3125 [hep-th]].
  %%CITATION = JHEPA,0807,005;%%
  \\
%\cite{Furuuchi:2008ki}
%\bibitem{Furuuchi:2008ki}
  K.~Furuuchi, S.~Y.~Shih and T.~Takimi,
  ``M-Theory Superalgebra From Multiple Membranes,''
  arXiv:0806.4044 [hep-th].
  %%CITATION = ARXIV:0806.4044;%%

  \bibitem{G08}
A.~Gustavsson,
  ``One-loop corrections to Bagger-Lambert theory,''
  arXiv:0805.4443 [hep-th].
  %%CITATION = ARXIV:0805.4443;%%

\bibitem{BLGM5}
P.~M.~Ho and Y.~Matsuo,
  ``M5 from M2,''
  JHEP {\bf 0806}, 105 (2008)
  [arXiv:0804.3629 [hep-th]].
  %%CITATION = JHEPA,0806,105;%%


%\cite{BLGM5-flux}
\bibitem{BLGM5-flux}
  P.~M.~Ho, Y.~Imamura, Y.~Matsuo and S.~Shiba,
  ``M5-brane in three-form flux and multiple M2-branes,''
  arXiv:0805.2898 [hep-th].
  %%CITATION = ARXIV:0805.2898;%%




\bibitem{LCGp-branes}
E.~Bergshoeff, E.~Sezgin, Y.~Tanii and P.~K.~Townsend,
  ``Super P-Branes As Gauge Theories Of Volume Preserving Diffeomorphisms,''
  Annals Phys.\  {\bf 199}, 340--365 (1990).
  %%CITATION = APNYA,199,340;%%

 \bibitem{LCGM2-SU(inf)}
J. Hoppe,  {\it Quantum Theory Of A Massless Relativistic Surface
And A Two Dimensional Bound State Problem},  PhD thesis, Massachuses
Institute of Technology, 1982, available at
http://www.aei.mpg.de/jh-cgi-bin/viewit.cgi
\\
B.~de Wit, J.~Hoppe and H.~Nicolai,
  ``On the quantum mechanics of supermembranes,''
  Nucl.\ Phys.\  B {\bf 305}, 545 (1988).
  %%CITATION = NUPHA,B305,545;%%


\bibitem{BST98}
E.~Bergshoeff, D.~P.~Sorokin and P.~K.~Townsend,
  {\it The M5-brane Hamiltonian},
  Nucl.\ Phys.\  {\bf B533}, 303 (1998)
  [arXiv:hep-th/9805065].
  %%CITATION = NUPHA,B533,303;%%

  %\cite{blnpst}
\bibitem{blnpst}
  I.~A.~Bandos, K.~Lechner, A.~Nurmagambetov, P.~Pasti, D.~P.~Sorokin and M.~Tonin,
  ``Covariant action for the super-five-brane of M-theory,''
  Phys.\ Rev.\ Lett.\  {\bf 78} (1997) 4332
  [arXiv:hep-th/9701149].
  %%CITATION = PRLTA,78,4332;%%


%\cite{schw5}
\bibitem{schw5}
M.~Aganagic, J.~Park, C.~Popescu and J.~H.~Schwarz,
  ``World-volume action of the M-theory five-brane'',
  Nucl.\ Phys.\  {\bf B496}, 191-214 (1997)
  [hep-th/9701166].
  %%CITATION = NUPHA,B496,191;%%

  %\cite{Pasti:1997gx}
\bibitem{Pasti:1997gx}
  P.~Pasti, D.~P.~Sorokin and M.~Tonin,
  ``Covariant action for a D = 11 five-brane with the chiral field,''
  Phys.\ Lett.\   {\bf B398} (1997) 41
  [arXiv:hep-th/9701037].
  %%CITATION = PHLTA,B398,41;%%

  \bibitem{CornM5mM2}
 J.~H.~Park and C.~Sochichiu,
  ``Single M5 to multiple M2: taking off the square root of Nambu-Goto
  action,''
  arXiv:0806.0335 [hep-th].
  %%CITATION = ARXIV:0806.0335;%%



  %\cite{Sorokin:1997ps}
\bibitem{Sorokin:1997ps}
  D.~P.~Sorokin and P.~K.~Townsend,
  ``M-theory superalgebra from the M-5-brane,''
  Phys.\ Lett.\  {\bf B412} (1997) 265
  [arXiv:hep-th/9708003].
  %%CITATION = PHLTA,B412,265;%%


%\cite{Carrollian}
\bibitem{Carrollian}
J. M. L\'evy-Leblond, ``Une nouvelle limite non-relativiste du
group de Poincar\'e'',  Ann. Inst. H. Poincar\'e 3 (1965) 1.

%\cite{Bacry:1968zf}
\bibitem{Bacry:1968zf}
  H.~Bacry and J.~Levy-Leblond,
  ``Possible kinematics'',
  J.\ Math.\ Phys.\  {\bf 9}, 1605 (1968).
  %%CITATION = JMAPA,9,1605;%%

%\cite{Gibbons:2003gb}
\bibitem{Gibbons:2003gb}
  G.~W.~Gibbons,
  ``Thoughts on tachyon cosmology,''
  Class.\ Quant.\ Grav.\  {\bf 20} (2003) S321
  [arXiv:hep-th/0301117].
  %%CITATION = CQGRD,20,S321;%%



\bibitem{mM2-f}
  I.~Jeon, J.~Kim, N.~Kim, S.~W.~Kim and J.~H.~Park,
  ``Classification of the BPS states in Bagger-Lambert Theory,''
  arXiv:0805.3236 [hep-th].
  %%CITATION = ARXIV:0805.3236;%%

\bibitem{Sagnotti+Mirian-Bonelli}
A.~Sagnotti and M.~Tsulaia,
  ``On higher spins and the tensionless limit of string theory,''
  Nucl.\ Phys.\   {\bf B682}, 83 (2004)
  [arXiv:hep-th/0311257];
  %%CITATION = NUPHA,B682,83;%%
\\
 G.~Bonelli,
  ``On the tensionless limit of bosonic strings, infinite symmetries and
  higher spins,''
  Nucl.\ Phys.\   {\bf B669}, 159 (2003)
  [arXiv:hep-th/0305155] and refs. therein.
  %%CITATION = NUPHA,B669,159;%%

  %\cite{KKKTP}
  \bibitem{KKKTP}
  P.~Claus, R.~Kallosh, J.~Kumar, P.~K.~Townsend and A.~Van Proeyen,
  ``Conformal theory of M2, D3, M5 and D1+D5 branes,''
  JHEP {\bf 9806}, 004 (1998)
  [arXiv:hep-th/9801206].
  %%CITATION = JHEPA,9806,004;%%



\end{thebibliography}
}
\end{document}